\documentclass[12pt]{article}
\usepackage[margin=1.0in]{geometry}
\usepackage[utf8]{inputenc}
\usepackage{amsthm,amsmath,physics,mathtools,amssymb}

\usepackage{charter}

\usepackage{CJKutf8}

\usepackage{enumitem}

\usepackage{hyperref}
\hypersetup{
    colorlinks=true,
    linkcolor=blue,
    filecolor=magenta,      
    urlcolor=cyan,
}
\usepackage[capitalise]{cleveref}

\theoremstyle{definition}
\newtheorem{theorem}{Theorem} 
\newtheorem{proposition}[theorem]{Proposition} 

\usepackage{graphicx}
\graphicspath{ {images/} }

\usepackage{tensor}

\DeclarePairedDelimiterX{\inp}[2]{\langle}{\rangle}{#1, #2}

\usepackage{xparse}

\NewDocumentCommand\LH{mo}{%
  \IfNoValueTF{#2}
   {\mathcal{L}(\mathcal{H}^{#1})}
   {\mathcal{L}(\mathcal{H}^{#1},\mathcal{H}^{#2})}%
}

\newcommand\id{\leavevmode\hbox{\small1\kern-3.3pt\normalsize1}}

\allowdisplaybreaks

\usepackage[T1]{fontenc}

\usepackage{bbold}

\usepackage{authblk}

\usepackage[title]{appendix}

\usepackage{mciteplus}

\usepackage{upgreek}

\usepackage{epigraph}

\usepackage{mdframed,lipsum}
\mdfdefinestyle{MyFrame}{%
    linecolor=black,
    outerlinewidth=2pt,
    innertopmargin=4pt,
    innerbottommargin=4pt,
    innerrightmargin=4pt,
    innerleftmargin=4pt,
        leftmargin = 4pt,
        rightmargin = 4pt
        }

\title{Path integral and particle ontology}

\author{Ding Jia (贾丁)\thanks{djia@perimeterinstitute.ca}}
\affil{Perimeter Institute for Theoretical Physics, Waterloo, Ontario, N2L 2Y5, Canada}
\affil{Department of Physics and Astronomy, University of Waterloo, Waterloo, Ontario, N2L 3G1, Canada}
\date{}

\begin{document}

\begin{CJK*}{UTF8}{gbsn}
\maketitle
\end{CJK*}

\begin{abstract}
It is often said that in relativistic quantum physics, a fundamental particle ontology is not viable. However, previous works have shown that certain relativistic quantum field theories can be reformulated as path integrals over particle configurations. Drawing on these works, I argue that a fundamental particle ontology is actually viable in relativistic quantum physics.
\end{abstract}

\section{Introduction}

\setlength{\epigraphwidth}{0.9
\textwidth}
\epigraph{[...] although it is not a theorem, it is widely believed that it is impossible to reconcile quantum mechanics and relativity, except in the context of a quantum field theory. A quantum field theory is a theory in which the fundamental ingredients are fields rather than particles; the particles are little bundles of energy in the field. There is an electron field, there is a photon field, and so on, one for each truly elementary particle.}{Weinberg, \textit{Towards the final laws of physics \cite{Weinberg1987TowardsPhysics}}}

Much of relativistic quantum physics is particle physics. Yet the theories behind particle physics are field theories. Is relativistic quantum physics fundamentally about fields or particles? 

Weinberg's quote reflects a popular view that quantum field theories (QFTs) are fundamentally about fields rather than particles. In related works on the foundations of relativistic quantum physics (e.g., \cite{Hegerfeldt1974RemarkLocalization, Redhead1982QuantumPhilosophers, Redhead1982QuantumPhilosophers, Bell1987BeablesTheory, Huggett1994WhatMatter, Teller1995AnTheory, Malament1996InParticles, Hegerfeldt1998CausalityEnergy, Hegerfeldt1998InstantaneousTheory, Weinberg1999WhatWas, Clifton2001AreTheory, WallaceEmergenceTheory, Halvorson2002NoTheories, 2002OntologicalTheory, Arageorgis2003FullingTheory, Colosi2008WhatParticle, Fraser2008TheInteractions, MacKinnon2008TheChallenge, Ruetsche2011InterpretingTheories, Struyve2011Pilot-waveTheory, Deckert2019ASea, Papageorgiou2019ImpactEntanglement, Feintzeig2021LocalizableTheory, Fraser2021ParticlesTheory, Jaeger2021TheFields, JiaWhatModel, SebensTheFields}), it is also an overwhelming majority view that a particle ontology cannot be maintained at the fundamental level.

Yet one thing appears puzzling. If a fundamental particle ontology was really so clearly impossible, why have we not stopped discussing it? Why is there such a long list of literature spanning decades all the way to the present on this topic?

Maybe it is because we want to understand why particle physics is ``particle'' physics. However, I believe there is a more basic reason that consciously or subconsciously drives us towards particles. A field is an entity that fills up the whole of spacetime. Yet our firsthand experience is about objects with delineated boundaries. Here is a table. Here is a chair. They do not fill up the whole of spacetime. They do not overlap. Our intuitions about these objects fit much more comfortably with a particle ontology than a field ontology. It is much easier to envision these objects as made of particles interacting with each other to form structured entities, than as excitations of fields filling out the whole universe. As such there is always the temptation to look for particles in the stories we tell about physics. 

Would it not be nice if after all there is a way to salvage a fundamental particle ontology?

In this work I argue that despite the contrary common wisdom, a fundamental particle ontology is viable in relativistic quantum physics. In previous literature, there have been both of perturbative and non-perturbative particle path integral reformulations of QFTs, some of which have in fact been available for decades. Their path integral configurations describe one-dimensional particle trajectories in spacetime, and can naturally be associated with a particle ontology. Because these particle path integrals are equivalent reformulations of standard QFTs, they do not suffer from the difficulties that alternative relativistic particle theories suffer. Therefore they show that a fundamental particle ontology is viable in relativistic quantum physics.

To my best knowledge the argument given here is missed in previous discussions on the foundations of relativistic quantum physics. I believe a major reason is that people focused too much on canonical and algebraic formulations, but did not pay enough attention to what new insights a path integral formulation can bring. Making a path integral consideration clarifies what is not necessary in a particle ontology in general contexts. As reviewed in \Cref{sec:csa} and \Cref{sec:ca}, many reasons have been considered before to rule out a fundamental particle ontology in relativistic quantum field theories, such as the lack of particle distinguishability, the lack of particle energy conditions, the lack of a fixed particle number, the lack of localizability, the lack of agreement on particle number among different observers, the lack of a unique Fock space representation in general curved spacetimes, the lack of applicability in interacting theories etc. While some of these may appear reasonable in canonical and/or algebraic formalisms, I will explain why they are unreasonable to demand in general in light of the particle path integral formalisms.

In summary, this work draws on the particle path integral formulations of QFT from previous works to argue for the viability of a fundamental particle ontology in relativistic quantum physics. The main message is \textit{viability}, and there are two points to be clear. First, the viability of a particle ontology in relativistic quantum physics itself does not imply that \textit{all} relativistic quantum theories are equipped with a particle ontology. Second, the viability of a particle ontology in relativistic quantum physics does not necessarily imply that it should be \textit{preferred} over other ontologies. As elaborated in  \cite{JiaWhatModel} and mentioned in \Cref{sec:pso}, while the scalar and fermionic parts of the Standard Model admit a particle interpretation, the gauge part can be naturally associated with a string interpretation. Therefore the QFT ontology debate is more sophisticated than just a particle vs. field debate. Other candidates, such as particle-string ontology, should also feature in the discussion. While preference question is an interesting topic, in this work I focus on the viability question since these are logically independent questions.

The rest of the work proceeds as follows. In \Cref{sec:pr}, I review some particle path integral reformulations of QFTs at both non-perturbative and perturbative levels. In \Cref{sec:csa}, I clarify in what sense these particle theories supply a particle ontology, and in which sense some addendum properties considered for particles are inessential. In \Cref{sec:ca}, I engage in some previous arguments against a fundamental particle ontology by offering counterarguments in light of the particle path integrals. In \Cref{sec:pso}, I briefly mention about the question whether a particle or a field ontology is preferable, and refer to a separate work to highlight particle-string ontology as an alternative for the Standard Model. In \Cref{sec:d}, I close with a discussion on the relevance of the ontological discussion to some open problems.

\section{Particle formulations}\label{sec:pr}

\subsection{What is special about path integrals?}\label{sec:mpi}

In the literature, foundational discussions of relativistic quantum physics often proceeds in canonical and algebraic formalisms. What motivates someone used to these formalisms to consider path integrals?

If one is really serious about the foundation of relativistic quantum physics for our universe, one should take the whole Standard Model into scope, instead of restricting to simple theories such as scalar field theories. While the canonical formalism is already clumsy to handle QED, it becomes unbearably clumsy to handle non-Ablian gauge theories.\footnote{Weinberg p.376 \cite{Weinberg1995TheFields}: ``The awkwardness in obtaining these simple results, which was bad enough for the theories considered in Chapters 7 and 8, becomes unbearable for more complicated theories, like the non-Abelian gauge theories to be discussed in Volume II, and also general relativity."} Similarly, the algebraic formalism in its current development is not so good at handling theories with interacting gauge and matter. In contrast, the path integral formalism is home to the Standard Model. It is \textit{the} language in which the Standard Model is taught and used \cite{Schwartz2013QuantumModel, Donoghue2013DynamicsEdition}.

\begin{figure}
    \centering
    \includegraphics[width=0.5\textwidth]{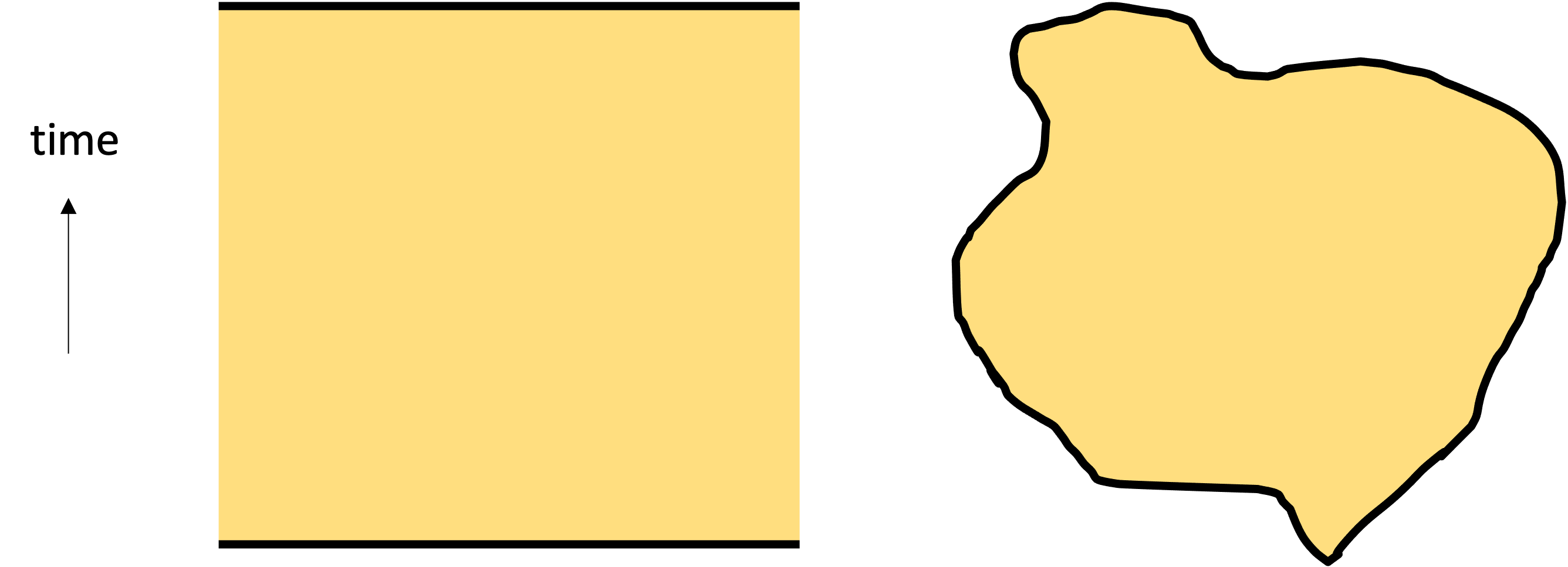}
    \caption{A path integral is applicable not only to spacetime regions with spatial boundaries (left), but also to spacetime regions of arbitrarily shaped boundaries (right).}
    \label{fig:region_pi}
\end{figure} 

The canonical formalism is clumsy, because its expressions carry a singled-out time. The path integral formalism is less clumsier, because it is a ``spacetime'' approach instead of a ``spatial'' approach. For an arbitrarily shaped spacetime region (\Cref{fig:region_pi}), the path integral partition function is given by
\begin{align}\label{eq:pf}
Z=\int D\Phi e^{iS[\Phi]},
\end{align}
where $S$ is the action of the theory, and the integral is over physical configurations $\Phi$ living in the spacetime region under consideration. It is unnecessary to single out a time direction to study the path integral. Moreover, It is unnecessary to restrict to spacetime regions with only spatial boundary components. This point is already noted in Dirac's 1933 seminal paper on path integrals \cite{Dirac1933TheMechanics}, and is developed to maturity by Oeckl in recent decades \cite{Oeckl2019APhysics}.

These have an important conceptual implication. On the one hand, the canonical formalism invites you to think of quantum states of spatial configurations evolving in time. It invites you to look for the ontology in the states. 
On the other hand, the path integral formalism as outlined above is not about spatial states evolving in time. The spacetime regions need not have spatial boundary components \cite{Colosi2008SpatiallyFormulation}, so spatial states are out of the window as a fundamental ingredient. There need not be a singled-out time foliation, so time evolution is out of the window as a fundamental ingredient as well. What then should one consider, if one is interested in ontological and foundational questions? If it is not about spatial states evolving in time, what is it all about? 


It is about physical configurations in spacetime. It is about boundary conditions. It is about interior conditions.

To see all these ingredients, one needs to go beyond the partition function  \eqref{eq:pf} and consider path integral formulas that yield probability amplitudes of the form
\begin{align}\label{eq:ppi}
Z[\psi,w]=\int D\Phi e^{iS[\Phi]} \psi[\Phi] w[\Phi].
\end{align}
The physical configurations $\Phi$ are the basic elements being integrated over. They live in spacetime rather than space. The boundary condition $\psi[\Phi]\in \mathbb{C}$ is a complex functional supported on the arbitrarily shaped boundary of the spacetime region to supply data such as the initial or final conditions of the universe. Here being supported on the boundary means $\psi[\Phi]=\psi[\Phi']$ when two spacetime configurations $\Phi$ and $\Phi'$ agree on the boundary. The interior condition $w[\Phi]\in \mathbb{C}$ is a complex functional supported within the spacetime region to supply data such as the measurement settings. 
For example, for a sharp measurement with two outcomes, there will be two corresponding sets $E_1$ and $E_2$ of physical configurations compatible with the outcomes. The interior conditions $w_i$ for $i=1,2$ are the characteristic functions so that $w_i[\Phi]=1$ when $\Phi\in E_i$ and $w_i[\Phi]=0$ otherwise. The theory predicts the probability ratio for the two outcomes as $\abs{Z[\psi,w_1]}^2/\abs{Z[\psi,w_2]}^2$.

Notably, the path integral formalism as presented above does not rely on Hilbert spaces. The physical configurations are functions over spacetime. The boundary and interior conditions are functionals on the physical configurations. None of them needs to be introduced as Hilbert space vectors or operators. For practical purposes, one could try to associate a Hilbert space structure to the set of boundary or interior conditions. However, this is not necessary in principle. The situation is similar to the algebraic formalism where Hilbert spaces can be constructed through the GNS construction but are not necessary to define the theories, and is in contrast to the canonical-operator formalism where states have to be introduced as Hilbert space elements, and time evolutions have to be introduced as operators. 

For someone interested in ontological and foundational questions, it is worthwhile to ask about the status of the interior conditions $w$. Can we be more specific the form it can take? Can we justify introducing it in some subregions of spacetime but not others, and where exactly should we introduce it? It is worthwhile to ask about the boundary condition $\psi$. How much can we know about it in our universe? Is it only possible to infer it by collecting data, or are there theoretical principles governing it? 

It is also worthwhile to ask about the physical configurations $\Phi$. What entities do they describe? What ontology does it suggest? This question will be the focus of the rest of this work. By the above discussion, it is hopefully clear that the ontology question takes a different form in the path integral formalism. Instead of focusing on physical configurations in space, one should look at physical configurations in spacetime. This means, for instance, if we think of a particle ontology, we should consider particle trajectories in spacetime instead of points in space.



\subsection{Non-perturbative formulation}\label{sec:npf}

We start by considering the ontological aspects of a real scalar field theory. As discussed at the end of this section, the particle ontology aspect of this theory carries over to complex scalar fields and fermionic matter fields of the Standard Model. Let us also start with a non-perturbative treatment of the theory, where we can spell out the definition of the path integral explicitly. The perturbative formulation is discussed in \Cref{sec:pf}.

\begin{figure}
    \centering
    \includegraphics[width=0.7\textwidth]{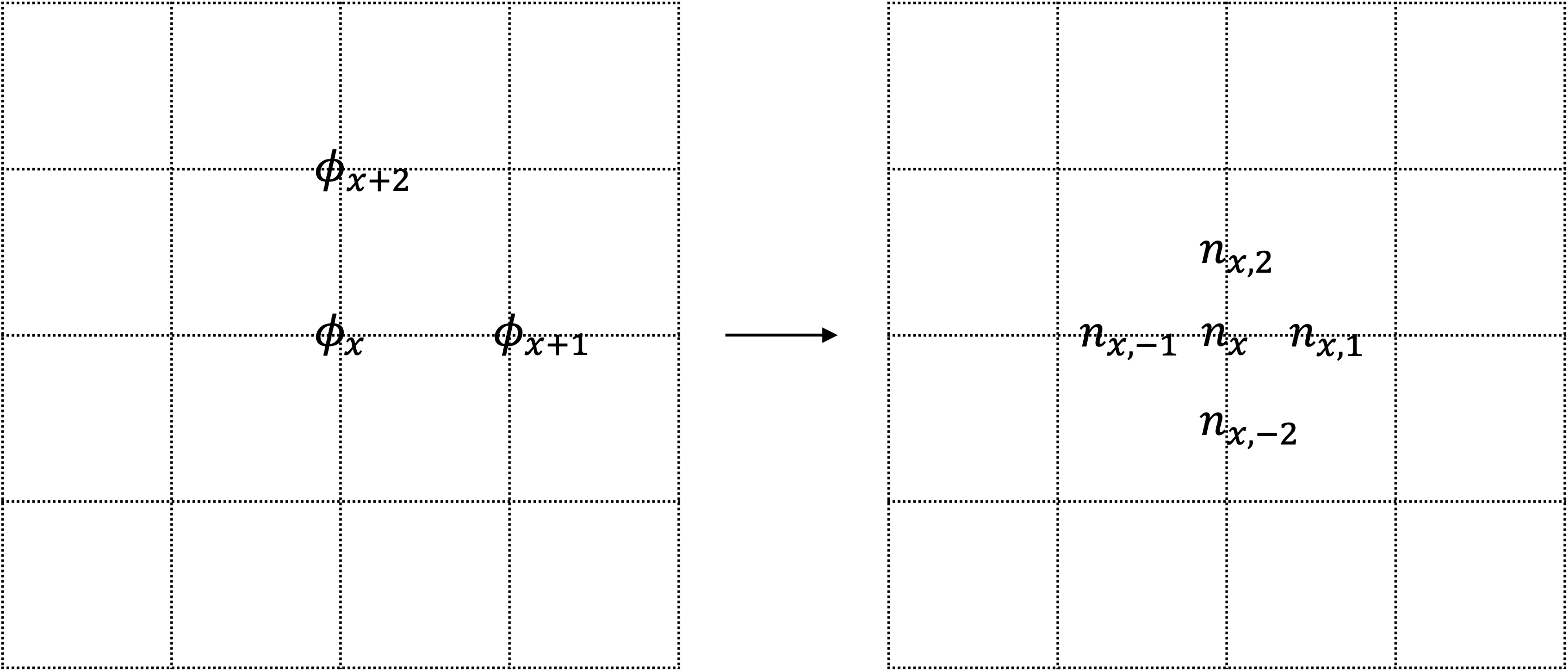}
    \caption{Lattice labelling conventions, with $\phi_x$ and $n_x$ at the vertices, and $n_{x,\nu}$ at the edges.}
    \label{fig:lattice-label}
\end{figure}

Consider a real scalar field theory in Minkowski spacetime with the Lagrangian density
\begin{align}
    \mathcal{L}=-\frac{1}{2}\partial^\nu \phi \partial_\nu \phi-\frac{1}{2} m^2\phi^2(x) - V(\phi)
\end{align}
where $V$ is a general potential, and the metric signature convention is $(-,+,+,\cdots)$.
To define the path integral non-perturbatively, the standard procedure is through a lattice \cite{Peskin1995}. Let there be a $D$-dimensional hypercubic lattice with spacing $a$ in both time and space directions. Rewriting derivatives as differences yields the lattice action
\begin{align}
S=&a^D \sum_x [ -\frac{1}{2} \sum_{\nu=1}^D g^{\nu\nu}(\frac{\phi_{x+\nu}-\phi_x}{a})^2 - \frac{1}{2} m^2 \phi_x^2 - V(\phi_x)]
\\
=& \sum_x [ \sum_{\nu=1}^D g^{\nu\nu} \tilde{\phi}_{x+\nu} \tilde{\phi}_{x} - \eta \tilde{\phi}_x^2  - \tilde{V}(\tilde{\phi}_x)].\label{eq:rsal3}
\end{align}
Here $g^{\nu\nu}$ is the Minkowski metric, $x$ refers to the lattice vertices, and $x\pm \nu$ refers to the vertex one unit in the positive or negative $\nu$-th direction away from $x$  (\Cref{fig:lattice-label}). In the last line we defined
\begin{align}
\tilde{\phi}_x &= a^{\frac{D-2}{2}}\phi_x,
\\
\eta &=a^2 m^2/2 + D-2,
\\
\tilde{V}(\tilde{\phi}_x)&=a^D V(\phi_{x}).
\end{align}
For simplicity, we will omit the tilde symbols in the following. The path integral partition function is given by
\begin{align}\label{eq:pfrsf}
    Z=\int D\phi e^{iS}.
\end{align}
Upon taking the lattice spacing limit $a\rightarrow 0$, this offers a non-perturbative definition of the quantum real scalar field theory.

To obtain the particle formulation we follow a procedure of Gattringer \textit{et al.} originally applied to avoid the sign problem in numerical simulations for Euclidean QFT \cite{Gattringer2013SpectroscopyGas, Gattringer2013LatticeField, Gattringer2016ApproachesTheory} (the Lorentzian formulation presented here was given in \cite{JiaWhatModel}). Let $S_{1}$ be the first term of (\ref{eq:rsal3}). With $\prod_{x,\nu}:= \prod_x \prod_{\nu=1}^D$ and $\sum_{n}:=\prod_{x,\nu} \sum_{n_{x,\nu}=0}^\infty$,
\begin{align}
e^{iS_{1}}=&\prod_{x,\nu} \exp{ i g^{\nu\nu} \phi_{x+\nu} \phi_{x} }
=\sum_{n}\prod_{x,\nu} \frac{( i g^{\nu\nu}\phi_{x+\nu} \phi_{x})^{n_{x,\nu}} }{n_{x,\nu}!} \label{eq:1l}
\\
=&\sum_{n}
(\prod_{x,\nu}\frac{( i g^{\nu\nu})^{n_{x,\nu}}}{n_{x,\nu}! })
(\prod_{x} \phi_x^{\sum_{\nu=1}^D(n_{x,\nu}+n_{x-\nu,\nu})}),
\\
Z=\int D\phi ~ e^{iS}=& \mathcal{N} \sum_{n} (\prod_{x,\nu} \frac{( i g^{\nu\nu})^{n_{x,\nu}}}{n_{x,\nu}!}) (\prod_{x} \int_{-\infty}^\infty d\phi_x ~\phi_x^{n_x} e^{-i \eta \phi_x^2-i V(\phi_x)}) \label{eq:rsfz}
\\
=& \mathcal{N} \sum_{n} (\prod_{x,\nu} \frac{( i g^{\nu\nu})^{n_{x,\nu}}}{n_{x,\nu}!}) (\prod_{x} f(n_x)),
\label{eq:rsfz1}
\end{align}
where $\mathcal{N}$ is a constant, $f$ stands for the last integral of \eqref{eq:rsfz}, and $n_x:=\sum_{\nu=\pm 1}^{\pm D} n_{x,\nu}$. 

Let us assume that the scalar field theories exhibit a global $Z_2$ symmetry so that $V(r)=V(-r)$. As explained in \cite{JiaWhatModel}, the general case without the $Z_2$ symmetry also admits a particle ontology. The difference is that the particle trajectories can break in the asymmetric case. On the other hand, the particle sector of the Standard Model obeys symmetries so that the particle trajectories do not break. Therefore we will assume $Z_2$ symmetry, under which
\begin{align}
f(n_x)=& \int_{-\infty}^\infty d\phi_x ~\phi_x^{n_x} e^{-i \eta \phi_x^2-i V(\phi_x)}
\\
=& \int_0^\infty d r_x ~ r_x^{n_x}e^{-i\eta r_x^2}[e^{-iV(r_x)}+(-1)^{n_x}e^{-iV(-r_x)}]
\\
=& \int_0^\infty d r_x ~ r_x^{n_x}e^{-i\eta r_x^2}[(1+(-1)^{n_x})e^{-iV(r_x)}]
\\
=& 2 \delta_2(n_x)\int_0^\infty d r_x ~ r_x^{n_x}e^{-i\eta r_x^2-iV(r_x)},
\\
Z=& \mathcal{N}\sum_{n}(\prod_{x,\nu} \frac{( i g^{\nu\nu})^{n_{x,\nu}}}{n_{x,\nu}!}) (\prod_{x} 2 \delta_2(n_x)\int_0^\infty d r_x ~ r_x^{n_x}e^{-i\eta r_x^2-iV(r_x)}).
\end{align}
In the second line the $\phi_x$ integral is separated into the positive and negative parts and recombined. In the third line, the $Z_2$ symmetry is used. In the fourth line $\delta_2(x)$ is the mod 2 Kronecker delta function, which arises because $1+(-1)^{n_x}=0$ for odd $n_x$.

The final result can be rewritten in a tidier form
\begin{align}\label{eq:fl}
Z=& \mathcal{N}\sum_{n\text{ extended}} \prod_{e} E_e(n_e) \prod_{x} V_x(n_x),
\end{align}
where $e$ relabels the lattice edges $(x,\nu)$,
\begin{align}
\sum_{n\text{ extended}}=&\sum_n \prod_{x} \delta_2(n_x),\label{eq:sne}
\\
E_e(n_e)=&\frac{( i g^{e})^{n_{e}}}{n_{e}!},
\\
V_x(n_x)=&2\int_0^\infty d r ~ r^{n_x}e^{-i\eta r^2-iV(r)}.
\end{align}
are the edge and vertex amplitudes.

\begin{figure}
    \centering
    \includegraphics[width=0.4\textwidth]{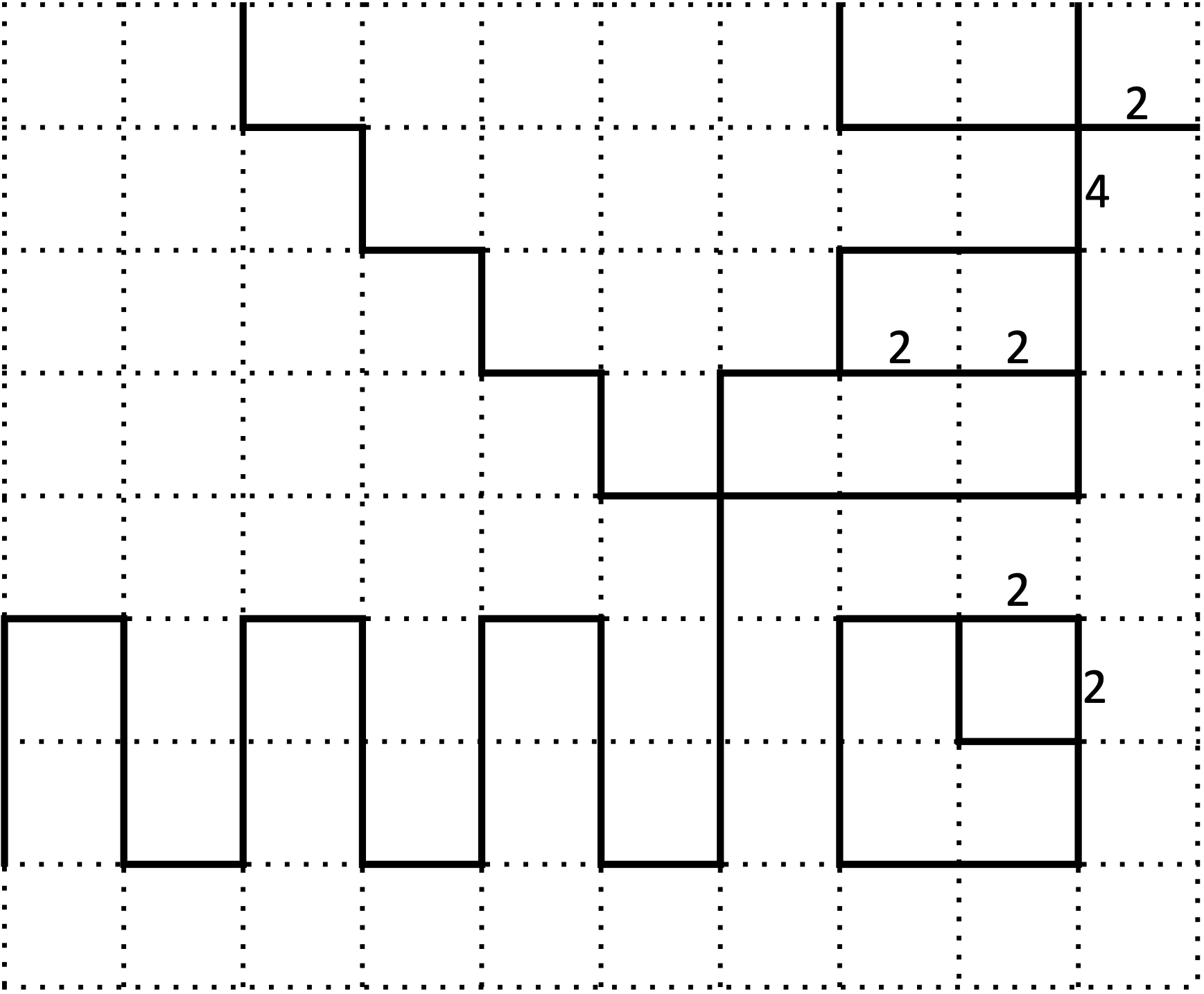}
    \caption{A lattice particle configuration on a finite region of 2D spacetime. Thickened edges are occupied by the labelled number of particle segment, with the label ``1'' omitted. One can check that the constraint $\delta_2(n_x)$ is obeyed in all bulk vertices, indicating that the particle trajectories keep extending until they close on themselves or reach the boundary of the spacetime region.}
    \label{fig:pc}
\end{figure} 

The above mathematical manipulations contain something physically profound. The steps from \eqref{eq:pfrsf} to \eqref{eq:fl} can be interpreted as shifting the ontology from fields to particles. We started with a path integral \eqref{eq:pfrsf} over field values $\phi$, but ended with a path integral \eqref{eq:fl} with $\sum_{n\text{ extended}}$. Here in \eqref{eq:sne}, $\sum_{n}=\prod_{x,\nu} \sum_{n_{x,\nu}=0}^\infty$, where the integers $n_{x,\nu}$ summed over arise in \eqref{eq:1l} from a Taylor series expansion for the kinetic term. Each $n_{x,\nu}$ is labelled by a vertex $x$ and a direction $\nu$, so we think of it as located to the edge $x,\nu$ connecting $x$ and $x+\nu$  (\Cref{fig:lattice-label}). What do these integers represent? 

The integer $n_{x,\nu}$ admits an interpretation as the number of particles passing the edge $x,\nu$. In this view, $n_x=\sum_{\nu=\pm 1}^{\pm D} n_{x,\nu}$ is the total number of particle line segments passing the vertex $x$. The presence of $\delta_2(n_x)$ in \eqref{eq:sne} then implies that the number of particles crossing any vertex is always even. This means that the segments of particles line that cross a vertex can always be grouped in pairs, so that the lines of particles continue to extend and never end. This explains the subscript ``extended'' in \eqref{eq:sne}. As a consequence, \eqref{eq:fl} can be interpreted as a path integral over extended particle configurations (\Cref{fig:pc}).

Note that the particle theory \eqref{eq:fl} and the field theory \eqref{eq:pfrsf} are mathematically equivalent. The gain of a particle theory comes at no cost other than the input of a quantum field theory. In addition, the above procedure to obtain a particle theory form a field theory is not limited to real scalar field theories, but also applies to complex scalar field theories and fermionic field theories \cite{JiaWhatModel}. Gauge theories do demand a slightly different procedure and call for a string ontology instead of a particle ontology. I will comment more on this aspect in \Cref{sec:pso}.

\subsection{Perturbative formulation}\label{sec:pf}

As discussed in \Cref{sec:csa}, the above non-perturbative particle formulations of QFT suffices to support a particle ontology. In the literature there are also perturbative particle formulations of QFT. For completeness, I review the gist of these formulations in this section.

The starting point is Feynman diagrams, which represent terms in a perturbation series expansion for the path integral amplitudes. Apart from the vertex factors and combinatorial factors, the non-trivial part of a Feynman diagram is the Feynman propagator. Traditionally, the Feynman propagators are obtained from a field theory. In the so-called ``worldline formalism'' \cite{EdwardsQuantumTheory}, the propagators are instead obtained from a relativistic particle path integral. This formalism has been applied not only in flat spacetime, but also in curved spacetimes \cite{Parker1979PathSpace, Bekenstein1981Path-integralSpacetimeb}. Here we consider the curved spacetime case for generality.

Consider a scalar field $\phi$ in a 4D curved spacetime whose classical field equation is $(\square+m^2+\xi R)\phi(x)=0$, where $m$ is the mass parameter, $\xi$ is the coupling constant, and $R$ is the Ricci scalar. In the so-called Schwinger proper time representation \cite{Schwinger1951OnPolarization} the Feynman propagator can be written as
\begin{align}\label{eq:spt}
G(x,y)=&i\int_0^\infty \braket{x,l}{y,0} e^{-im^2 l} dl,
\end{align}
where $\braket{x,l}{y,0}$ is the relativistic particle path integral defined by
\begin{align}
\label{eq:wlpi}
\braket{x,l}{y,0}=&\int D[x(l')] \exp{i\int_0^l dl' [\frac{1}{4}g_{ab}\frac{dx^a}{dl'}\frac{dx^b}{dl'}-(\xi-\frac{1}{3})R(l')]}
\\
:=&\lim_{N\rightarrow \infty}\Big[\frac{1}{i}(\frac{1}{4\pi i \epsilon})^2\Big]^{N+1}\int \prod_{n=1}^N d^4 x_n [-g(x_n)]^{1/2} 
\nonumber\\
& \exp{\sum_{j=0}^N i\int_{j\epsilon}^{(j+1)\epsilon}[\frac{1}{4}g_{ab}\frac{dx^a}{dl'}\frac{dx^b}{dl'}-(\xi-\frac{1}{3})R(l')] dl' }.\label{eq:mac}
\end{align}
Here the path integral is over relativistic particle paths starting at $x_0=x$ and ending at $x_{N+1}=y$. As usual, the path integral is defined over piecewise linear paths sequentially connecting the points $x_n$ for $n=0,\cdots,N$ in spacetime. The paths are relativistic because all 4 components of $x_n$ are integrated over, unlike the non-relativistic case where only the 3 spatial components are integrated over. Instead of the coordinate time in the non-relativistic case, the paths are parametrised by the so-called Schwinger proper time $l'$, which starts at $0$ and ends at $l$, so that the parameter spends $\epsilon:=l/(N+1)$ on each piecewise linear segment of a path. Between the two ends of a segment, the action is evaluated along a geodesic that connects the points, with $g_{ab}$ as the metric field and $g$ as its determinant. The arbitrary value $l$ should not feature in the final result, so in \eqref{eq:spt} it is integrated over. Finally, the term $m^2$ is taken to contain an infinitesimal negative imaginary part according to the Feynman prescription.

\begin{figure}
    \centering
    \includegraphics[width=0.8\textwidth]{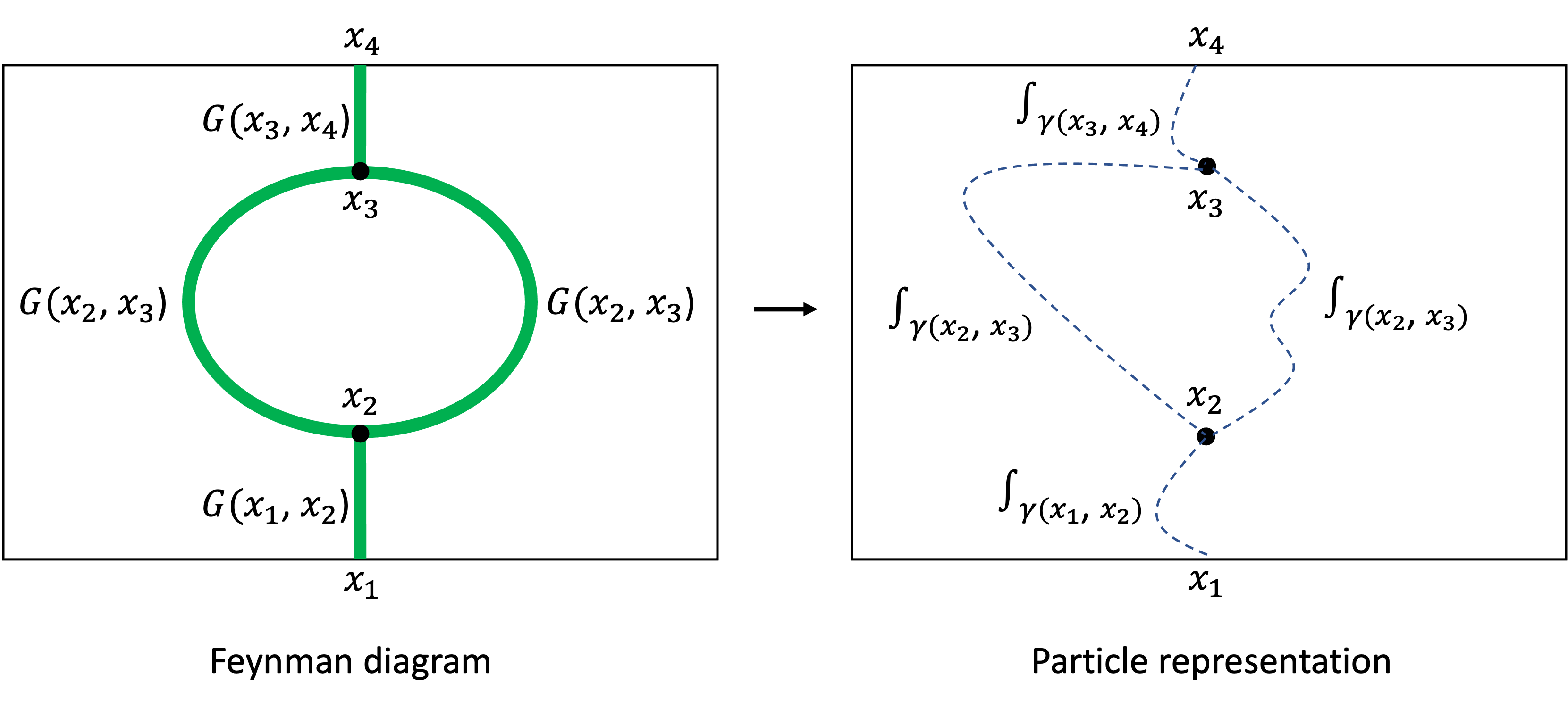}
    \caption{An ordinary position space Feynman diagram (left) represented as a particle path integral (right). Each Feynman propagator $G(x,y)$ is represented as a sum over particle trajectories $\gamma(x,y)$ with fixed end points $x$ and $y$.}
    \label{fig:pr}
\end{figure} 

In this formalism, a position space Feynman diagrams can be thought of as a particle path integral, with each Feynman propagator representing the sum over a class of relativistic particle paths connecting its two vertices (\Cref{fig:pr}). 
One could equip theories in this formalism with a particle ontology along this line of reasoning. In the following, I will focus on the non-perturbatively defined theories to discuss particle ontology instead. This avoids the reliance on Feynman diagrams and perturbation series.

\section{Clarifying the particle ontology}\label{sec:csa}

A particle ontology can be naturally attributed to the theories reviewed in \Cref{sec:npf}. As discussed in \Cref{sec:mpi}, in a path integral theory the ontology is sought for in the spacetime physical configurations. In the particle path integrals, the configurations are specified by one-dimensional extended lines in spacetime, which are naturally interpreted as particle trajectories. 

This supports a particle ontology because particles trace out one-dimensional trajectories in spacetime, and conversely whatever that trace out one-dimensional trajectories in spacetime are reasonably referred to as particles. However, in the literature other attributes for a particle ontology have been discussed, and these attributes need not hold in the particle path integral theories. This should not jeopardize the particle ontology for the particle path integral theories, not only because they do not obstruct the above reason for a particle ontology based on one-dimensional trajectories, but also because it is hard to conceive of any better ontology for these particle path integral theories. Nevertheless, to avoid confusion I shall point out in this section what other attributes do not apply in the particle path integrals, and explain further why they do not obstruct a particle ontology.



\subsection{Distinguishability}\label{sec:dy}

The particles in the path integral configurations are not distinguishable. For instance, the boundary conditions are specified by particle numbers without naming the particles. The boundary condition can be matched by any particle configuration, as long as the particle numbers are matched.

One possible requirement for particles is that they must be capable of bearing labels, so they must be distinguishable. Although I am not aware of any researcher who insists on this requirement, it has been mentioned in discussions of particle ontology \cite{Teller1995AnTheory, Ruetsche2011InterpretingTheories, Fraser2021ParticlesTheory}.

Even in non-relativistic statistical mechanics, we routinely talk about identical ``particles'' and their statistical distributions. To the extent that the notion of a particle is meaningful in statistical mechanics, there is no reason for objecting to a particle ontology for relativistic quantum physics based on indistinguishability.

\subsection{Energy-momentum relation}\label{sec:ec}

In Fraser \cite{Fraser2008TheInteractions} a relativistic energy-momentum relation is required to hold for ``quanta'', a term used by Teller \cite{Teller1995AnTheory} to refer to particles that cannot bear labels. For instance, in the Fock space for a free theory, the one-particle states
\begin{align}
a^\dagger(\textbf{k},t)\ket{0}
\end{align}
obey the relativistic energy-momentum relation $E=\sqrt{\textbf{k}^2+m^2}$.

This requirement may be reasonably considered for Hilbert space-based theories, but is not so reasonable in the context of path integrals. Even for the particle configurations of a non-relativistic particle path integral, it is not reasonable to demand the non-relativistic energy-momentum relation. This is because in a path integral the particle trajectories can be ``off-shell'', i.e., they do not obey the classical equation of motion, and the non-relativistic energy-momentum relation does not hold for them. Unless one objects a particle ontology already for the non-relativistic path integral theories, there is no reason to object a particle ontology for relativistic theories based on energy-momentum relation.

\subsection{Fixed-number}\label{sec:fn}

In the same event that Weinberg delivered the quote at the beginning, Feynman gave a talk to explain the necessity of antiparticles in relativistic quantum physics \cite{Feynman1987TheAntiparticles}. This gives a very intuitive explanation why a relativistic quantum theory of particle has to involve a variable number of particles, so we review it (in an adapted form) here. 

\begin{figure}
    \centering
    \includegraphics[width=0.8\textwidth]{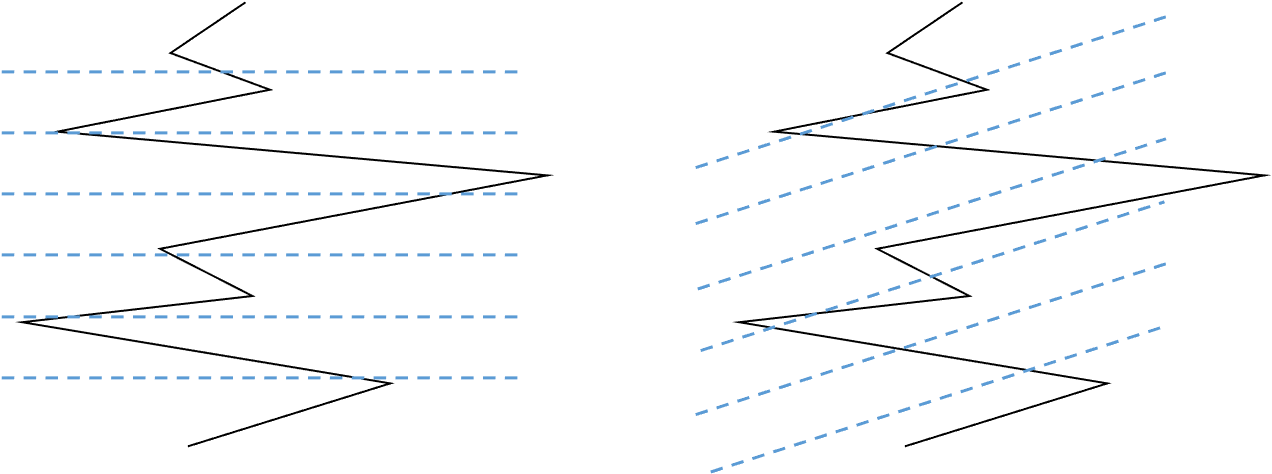}
    \caption{A same particle path integral configuration in relativistic spacetime under two time foliations. On the left, the path crosses each spatial slice once. On the right, the path crosses the spatial slices a variable number of times.}
    \label{fig:pi_particle}
\end{figure}

Consider a path integral configuration for ``one'' particle, such as the one shown in \Cref{fig:pi_particle}. Such zigzagging configurations are familiar from non-relativistic particle path integrals, but we consider them in a relativistic spacetime. In the ``horizontal'' time foliation (left of \Cref{fig:pi_particle}), the path crosses each spatial slice once as if the path were for one particle. However, in another foliation (right of \Cref{fig:pi_particle}) the path can cross the spatial slices a variable number of times. This property holds for generic particle path integral configurations. In this sense, the path integral is actually for a variable number of particles instead of one particle. 


Such path integral configurations with a variable number of particles feature in the theories of \Cref{sec:pr}. For the perturbative theories of \Cref{sec:pf}, the number of particles varies because the time component of the particle locations $x_n$ is integrated over, so the particle also ``zigzags'' in time. For the non-perturbative theories of \Cref{sec:npf}, the number of particles varies because all extended particle line configurations are summed over.

In some discussions about particle ontology, it is mentioned in passing as a possible requirement that the particle number remains fixed \cite{Ruetsche2011InterpretingTheories, Fraser2021ParticlesTheory}. While one may reasonably draw a distinction between theories where particle numbers can change and those where they cannot, I find it unreasonable to forbid the latter theories from possessing a particle ontology. Path integral configurations like in \Cref{fig:pi_particle} show up in both non-relativistic and relativistic theories. It would be very confusing to say that in one case they describe particles while in the other case they do not. A better option is to accept that the configurations are particle configurations also in relativistic theories, and allow particle numbers to change.

\section{Counterarguments}\label{sec:ca}

In a very nice recent review, Fraser \cite{Fraser2021ParticlesTheory} collected four arguments against particle ontology in relativistic quantum physics from previous literature. In this section I address these negative arguments and argue that none of them obstructs the particle ontology of the path integral theories of \Cref{sec:pr}.

\subsection{Localizability}



In an influential paper entitled ``Why There Cannot be a Relativistic Quantum Mechanics of (Localizable) Particles'' \cite{Malament1996InParticles} (see also \cite{Halvorson2002NoTheories}), Malament proved that there cannot be a meaningful particle position operator in relativistic quantum theories. Some subsequent discussions about particle ontology focus on localizable particles that come with position operators (e.g., \cite{Papageorgiou2019ImpactEntanglement}), while others do not (e.g., \cite{Fraser2008TheInteractions}).

For reasons mentioned at the end of \Cref{sec:fn}, the particle number should be allowed to change while maintaining a particle ontology. In the path integral language, a basis for the boundary conditions on a spatial slice includes both one particle and multiple particle configurations. We cannot expect that a position operator with a single real eigenvalue to meaningfully reflect the positions of the multiple particles. The lack of a meaningful position operator is an important finding, but this should not be misread as excluding particles as fundamental entities in relativistic quantum physics. 
Instead, a non-fixed number of particles is allowed in a relativistic quantum theory, and the lack of a meaningful position operator is a natural consequence of allowing the particle number to change.

\subsection{Unruh effect}

The Fulling–Davies–Unruh effect \cite{Fulling1973NonuniquenessSpace-Time, Davies1975ScalarMetrics, unruh_notes_1976}, or Unruh effect in short, shows that in QFTs, observers travelling in the same universe on different trajectories need not see the same number of particles. 

One might naively think that this undermines a particle ontology. However, the Unruh effect has been reproduced in the particle path integral formulation of QFTs, for example by Rosabal \cite{Rosabal2018NewEffect, Rosabal2020EverythingEffectb}. For all observers, the path integral configurations are still particle configurations. The difference in the detected particle number spectrum is due to the different boundary conditions imposed for the different observer trajectories. The argument in support of the particle ontology given in \Cref{sec:csa} is based only on the fact that the path integral configurations are specified as particle configurations. The Unruh effect does not change this, so it does not undermine a particle ontology.

\subsection{Curved spacetime}

Citing Wald \cite{Wald1994QuantumThermodynamics}, Fraser points out the lack of a unique Fock space representation in general curved spacetimes as another potential argument against particle ontology. Regardless of whether this negative argument is valid in the context of a particle interpretation based on Fock spaces, it does not apply to the present particle interpretation based on path integrals. As pointed out in \Cref{sec:mpi}, the path integral formalism does not rely on Hilbert spaces.

The perturbative particle path integrals of \Cref{sec:pf} are directly applicable to curved spacetimes. The non-perturbatively particle path integrals of \Cref{sec:npf} have so far been applied to flat spacetime, but there appears to be no obstacle in generalising it to curved spacetimes (e.g., using simplicial lattices \cite{Brower2016QuantumTheory}). Just like in flat spacetime, the path integral configurations are specified by particle configurations. The only difference is that now the trajectories live in curved spacetimes. This difference does not change anything about the arguments for a particle ontology given in \Cref{sec:csa} based on one-dimensional trajectories in spacetime but without reference to Fock spaces. 

\subsection{Interaction}

In \cite{Fraser2008TheInteractions} Fraser points out that the usual ``quanta'' interpretation for QFT based on the Fock space of free systems is not viable once interactions are introduced. This may be read as an argument against particle ontology. As recognized by Fraser \cite{Fraser2021ParticlesTheory}:
\begin{quote}
Of course, a particle ontology could be restored by finding a way to give a particle interpretation of QFT that does not rely on Fock space, but this is not a plausible option until there is a concrete proposal for how this interpretation would proceed.
\end{quote}

The intention of this work is precisely to give a concrete proposal for a particle interpretation of (some) QFTs through the path integral theories of \Cref{sec:pr} without relying on Fock spaces, or any Hilbert spaces (\Cref{sec:mpi}). With or without interactions, the path integral configurations of \Cref{sec:npf} in spacetime are always specified by the particle configurations. Therefore turning on interactions does not affect the particle ontology.

\section{Particle-string ontology for the Standard Model}\label{sec:pso}

The viability of a particle ontology naturally raises a question. Is there a preference between a particle ontology and a field ontology? This question is discussed, for instance, by Sebens recently in \cite{SebensTheFields}, with an answer in favour of a field ontology. 


As far as the Standard Model goes, my view given in \cite{JiaWhatModel} is to favour a third alternative, the particle-string ontology. The standard model contains scalar, fermion, as well as gauge degrees of freedom. While the scalar and fermion parts admit particle representations, the gauge part is most comfortably represented as strings in a non-perturbative setting. This fits well with Faraday's old idea that the interactions among particles are mediated by extended lines \cite{Faraday1852OnForce}. 

The difficulty of a particle ontology in incorporating states for photons is highlighted by Sebens \cite{SebensTheFields}. The ease of incorporating gauge degrees of freedom in a particle-string ontology is one reason to prefer it over a particle ontology. In \cite{JiaWhatModel} I argue that a particle-string ontology is also preferred over a field ontology since the particle-string picture offers a very intuitive explanation for gauge symmetries of the Standard Model as the feature that particles and open strings are always coupled. In addition, the particle-string path integral is more economic because it excludes from the outset the redundant path integral configurations present in a field formulation. Yet the reasons for a field ontology against a particle ontology mentioned in \cite{SebensTheFields} are worth considering further in comparison to a particle-string ontology, and I view it as a still ongoing discussion whether for the Standard Model a particle-string ontology is truly preferable over other ontologies.

\section{Discussion}\label{sec:d}

Up to this point the discussions are situated in the context of QFT in classical spacetime, which is a well-understood topic. On the other hand, the topics of relativistic particle path integrals and particle ontology are also relevant to some less-understood open questions of physics. I finish with a very brief discussion related to quantum gravity and measurements.

In quantum gravity, most candidate theories we have (simplicial quantum gravity, dynamical triangulations, spinfoams, group field theories, quantum causal sets, effective field theories, functional renormalization group theories, and Polyakov string theory etc.) are path integral theories \cite{oriti2009approaches}. In these theories, matter coupling proceeds naturally through a joint gravity-matter path integral\footnote{Or a unified path integral for strings in string theory.}. In discussions on the ontology of truly fundamental relativistic quantum physics that incorporates quantum spacetime, path integrals cannot be missed.

Describing measurements in relativistic quantum physics is a thorny issue \cite{Sorkin1993ImpossibleFields, Beckman2002MeasurabilityOperators, PagePossibleBeings} that is receiving renewed attention. The recent discussions of \cite{Borsten2021ImpossibleRevisited, Jubb2022CausalTheory, Bostelmann2021ImpossibleApparatus} are based in the canonical and algebraic formalisms. It is imperative to develop a path integral description of measurements. This description needs to refer to the path integral configurations. Here the particle(-string) formulation offers a helpful alternative to the field formulation. It may be especially helpful in gauge theories since the particle-string formulation are written directly in terms of gauge invariant quantities.

\section*{Acknowledgement}

I am very grateful to Chip Sebens for discussions that helped improved an earlier draft, and to Lucien Hardy and Achim Kempf for long-term encouragement and support. Research at Perimeter Institute is supported in part by the Government of Canada through the Department of Innovation, Science and Economic Development Canada and by the Province of Ontario through the Ministry of Economic Development, Job Creation and Trade. 

\bibliographystyle{unsrt}
\bibliography{references.bib}

\end{document}